# Compression stress in opposite wood of angiosperms: observations in chestnut, mani and poplar

Bruno CLAIR [a, b, *]

Tancrède ALMERAS [c]

Junji SUGIYAMA [a]

[a] Laboratory of Biomass Morphogenesis and Information, Research Institute for Sustainable Humanosphere, Kyoto University, Uji-Kyoto 611-0011, Japan

[b] Laboratoire de Mécanique et Génie Civil (LMGC), UMR 5508 CNRS – Université Montpellier 2, Place E. Bataillon, CC 48, 34095 Montpellier CDX 5, France

[c] Laboratory of Bio-Material Physics, Graduate School of Bioagricultural Sciences, Nagoya University, Nagoya 464-8601, Japan

[*] Tel: (33) 4 6714 4918 - fax: (33) 4 6714 4792 - e-mail : clair@lmgc.univ-montp2.fr

RÉSUMÉ

Pour faire face aux contraintes environnementales, les arbres sont capables de contrôler l'état de contraintes des nouvelles couches de bois formé pour réorienter leurs axes. Angiospermes et gymnospermes ont évolués vers deux stratégies différentes : les premiers produisent un bois à fortes précontrainte de tension sur la face supérieure de l'axe incliné alors que les derniers mettent en place un bois en précontrainte de compression sur la face inférieure. La différence de contrainte avec la face opposée constituée de bois en légère tension produit un même résultat, la flexion de l'axe. Pourtant, au hasard des expérimentations, des valeurs de compression ont plusieurs fois été observées dans le bois opposé des angiospermes. Cette étude reprend des donnés anciennes sur le châtaignier et le manil et les complète avec des données sur peuplier pour mettre en évidence qu'il ne s'agit par d'erreurs de mesure mais réellement d'une stratégie de ces arbres consistant à produire du bois en légère précontrainte de compression pour une meilleure efficacité du redressement.

SUMMARY

In order to face environmental constraints, trees are able to re-orient their axes by controlling the stress level in the newly formed wood layers. Angiosperms and gymnosperms evolved in two distinct strategies: the former produce a wood with large tension pre-stress on the upper side of the tilted axis, while the latter produce a wood with large compression pre-stress on the lower side. In both cases, the difference between this stress level and that of the opposite side, in light tension, generates the bending of the axis. However, light values of compression were sometimes measured in the opposite side of angiosperms. By analysing old data on chestnut and mani and new data on poplar, this study shows that these values were not measurement artefacts. This reveals that generating light compression stress in opposite wood is a strategy for optimising the biomechanical efficiency of the re-orientation mechanism.

1. **INTRODUCTION**

During their development, trees are submitted to various mechanical constraints that may cause a permanent change in the orientation of their stem and branches. These constraints are environmental, such as soil instability or damages due to the action of wind or snow, as well as biological, such as the weight of epiphytes or the weight of the stem and crown of the tree itself. Controlling and correcting the orientation of woody axes is then necessary for a tree to adapt to its environment, ensure its long-term mechanical stability and maximise its chances to survive and reproduce [15].

This morphological plasticity is achieved through the production of reaction wood on one side of the stem. Reaction wood is characterised by the high magnitude of the longitudinal stress generated inside wood cells during maturation. The asymmetric distribution of maturation stress within the newly formed layers of a stem cross-section creates a bending moment resulting in a change in stem curvature [2, 3, 14]. Regarding this mechanism of stem re-orientation, gymnosperm and angiosperm species have different strategies. Gymnosperms produce compression wood in the lower side of the tilted stem and angiosperms produce tension wood in the upper side [3, 5, 14, 20, 24, 26]. In both cases the opposite side is characterised by a "normal" stress, i.e. a light tension.

Because of the stiffness of the internal wood layers, most of the maturation stress in the outer layers of the tree is not released *in situ* [3]. This can be evidenced and quantified by releasing the longitudinal stress at the surface of wood and measuring the resulting strain, referred to as residual growth strain (RGS). Typical RGS values range from –0.1% to –0.4% for tension wood (TW), from –0.015% to –0.1% for normal wood (NW) and from +0.05% to +0.3% for compression wood (CW) [14].

Positive RGS values were sometimes measured in angiosperm opposite wood [10, 11]. These values were believed to be measurement artefacts. Indeed, metrological studies showed that the methods for measuring RGS may have a non-negligible incertitude [14, 29]. Also, other sources of mechanical stress are a possible cause of error for the estimation of maturation stress. Any increment of load or change in stem orientation generates an additional stress field inside the tree, referred to as support stress. Growth stress is the sum of maturation stress and support stress. It can be shown that the support stress is low in peripheral wood layers, and generally much lower than maturation stress of reaction wood [3]. However, in certain conditions, it may not be negligible so that the measurement of RGS, even if accurate, is a biased estimation of the maturation strain. This bias may not be negligible when compared to the magnitude of RGS in normal wood.

Recently, performing research on tension wood of species well known to produce high tension stress, some positive RGS value was measured on the opposite side (lower side) of the tilted trees [10, 11]. In this paper, by considering other physical and micro-structural features usually correlated to the RGS value, we aim at verifying if these measurements were artefacts or if angiosperms develop compression maturation stress in opposite wood to optimize their biomechanical efficiency.

## 2. MATERIAL AND METHODS

Studies were performed on tilted trees of chestnut (*Castanea sativa* Mill.), mani (*Symphonia globulifera* L.f.) and poplar (*Populus euramericana* Guinier). Diameters at breast height were respectively 15 cm, 20 cm and 25 cm. Chestnuts were growing in a private forest in south of France, mani in a tropical rain forest in French Guyana (South America) and poplar in Kyoto university experimental forest (Japan). Experiments on mani and chestnut involved measurements of RGS and longitudinal drying shrinkage, and were presented in details in previous papers [10, 11]. New data on poplar are also shown, including measurements of RGS and micro-fibril angle.

## 2.1 Residual growth strain (RGS)

Various methods are available for measuring RGS. They are based on the same principle, which consists in measuring the longitudinal strain at the wood surface, after artificially releasing the stress [14, 29]. For chestnuts and mani, the "single hole" method was used [11, 14]. With this method, a displacement is measured (in μm) and converted into a strain (in %) using a calibration factor. Empirical and theoretical calibration factors were provided by Fournier *et al.* [14]. We used the factor $12.3 \times 10^{-4}$ corresponding to a standard hardwood. For poplar experiments, RGS was directly obtained using the strain gauge method described by Yoshida and Okuyama [29].

In the three species, RGS were measured at various positions around the circumference. For each RGS measurement, several measurements of the other feature were performed, at various locations close to the point where RGS were measured.

## 2.2 Longitudinal shrinkage

The length of chestnuts sample ($R \times T \times L = 1 \times 5 \times 50$ mm) and the thickness of a mani disk ($L = 50$ mm) were measured in green and oven-dried condition with a Mitutoyo transducer. Shrinkage was calculated as the dimensional variation between green and dry states, divided by the dimension in the green state [10, 11].

## 2.3 Microfibrils angle (MFA)

MFAs were measured in green conditions. A plain sawn plate ($R \times T \times L = 2 \times 0.5 \times 20$ mm) was mounted in a fibre goniometer, keeping $\theta$-$2\theta$ to satisfy the Bragg condition of (004) meridional planes. The intensity distribution was then recorded by rotating the sample ($\beta$ rotation) with a speed of 6 degrees per minute. The CuK$_\alpha$ ($\lambda$=0.1542 nm) were generated by X-ray generator (RIGAKU Ultra18HF) operated at 30kV, 300mA. Parameter T defined by Cave [7] was graphically derived from the diffraction intensity diagrams. The average MFA of each sample was estimated using Cave's formula: MFA = 0.6×T. As stated by Cave [8], the intensity diagram obtained from the (004) plane is identical to that obtained from the (200) plane, but suffers contamination from other weak crystalline planes having a similar Bragg angles. Considering that this contamination is low, parameter T gives useful estimations of the MFA. Also, application of Cave's formula for wood samples having a MFA lower than 20° is subject to discussion. Values obtained using the "improved Cave's method" [27] were also computed to check the consistency of the results.

## 3. RESULTS AND DISCUSSION

*Figure 1* and *2* show the longitudinal shrinkage measured on chestnut and mani versus the RGS. Results on tension wood are presented in order to show the ability of these trees to produce high tensile stressed wood, having a high longitudinal shrinkage, as it is well known [9, 13, 17, 22]. Observation of the measured longitudinal shrinkage versus RGS shows a continuum in behaviour from the "normally" tensile stressed wood to the compression stressed wood. Inside the group of normal wood samples (including compression values), RGS and longitudinal shrinkage are significantly correlated (R=0.60, p<0.001 for chestnut, R=0.59, p<0.05 for mani). Moreover, the mean longitudinal shrinkage of samples with positive RGS is significantly higher than that of normal wood samples having negative RGS (Mann-Whitney U test, p<0.01 for chestnut, p<0.05 for mani). This tendency is similar to that observed in gymnosperms in the transition between normal and compression wood [23, 25].

*Figure 3* shows the relation between MFA estimation and RGS for poplar. Results obtained on tension wood are presented in order to confirm the ability of this tree to generate high tension stresses associated with a very low MFA[1], as typically found in gelatinous fibres [16]. For samples with RGS higher than -0.1%, there is a general trend for the MFA to increase with the RGS. Inside the group or normal wood samples (including compression values), RGS and MFA are significantly correlated (R=0.67, p<0.005). The mean MFA of samples with positive RGS is significantly higher than that of normal wood samples having negative RGS (Mann-Whitney U test, p<0.001). Again, this tendency is typical of the relation observed in gymnosperms from normal wood to compression wood [18, 25].

On the three angiosperms species studied, positive values of RGS between +0.01% and +0.02% were found. Each time, the values of other features positively correlated to the RGS were found consistently higher in samples with positive RGS. This clearly confirms that the observed positive RGS are not due to measurement errors, but really correspond to a compressive maturation stress.

In the case of poplar, our measurements show that wood with compression maturation stress have higher MFA than typical normal wood. Unfortunately, the MFA of chestnut and mani samples were not measured. However, several experimental and theoretical works showed that the increase in longitudinal shrinkage from normal to compression wood is due to the increase of the MFA [18, 25], so that it is likely that a similar observation would have been done on these species. Several works showed that development of compressive growth stress is a direct mechanical consequence of this large MFA [1, 6, 28].

---

[1] Values around 5° found for tension wood MFA should not be considered as a correct estimate, since the Cave's method [7] used to derive the MFA from the diffraction diagram is not adapted to such low values. Using Yamamoto's improved method [27], negative values ranging between -1° and -5° were obtained. The real mean value of MFA in tension wood is probably close to 0°.

Even if the RGS value of this wood is lower than that of typical CW of gymnosperms, similar features were observed, concerning the cell structure (MFA), the physical behaviour (shrinkage) and the biomechanical behaviour (RGS). For all features, this opposite wood of angiosperm seems to be an intermediate form between normal wood and typical compression wood, maybe similar to what was observed in gymnosperms in the transition zone between these two tissues [30].

The presence of CW in angiosperms was previously reported, to our knowledge, only for the case of *Buxus* (*microphylla* and *sempervirens*) [4, 20, 31] and other primitive angiosperms such as *Pseudowintera colorata* (Winteraceae) [19]. In the case of these species, the magnitude of RGS, anatomical observations and/or chemical analyses showed that typical CW was produced close to CW of gymnosperms [4, 19, 31]. This particular case clearly differs from the results of our study from a biomechanical point of view. Indeed, the biomechanical mechanism of *Buxus* and *Pseudowintera* is completely similar to that of gymnosperms, producing CW with stem eccentricity in the lower side associated and NW in the opposite side.

Trees studied in this paper seem to cumulate advantages of angiosperms and gymnosperms by combining the production of strong TW on one side and a kind of light CW on the opposite side.

Values of RGS in tension wood measured for many species suggest that there may be an upper limit in the tensile stress that can be generated in wood cell walls [12, 21]. In this case, the biomechanical efficiency of the re-orientation process can only be improved through complementary mechanisms. Difference in stiffness between NW and TW or stem eccentric growth are known examples of such complementary mechanisms [2]. The simultaneous production of TW on the upper side and CW in the lower side is an additional mechanism, which allows a substantial improvement of the biomechanical efficiency. Indeed, the change in curvature induced by reaction wood production is roughly proportional to the difference in RGS between the 2 sides. For our samples, this difference was increased by 17% to 38% because light compression wood was produced on the opposite side instead of typical normal wood.

## 4. CONCLUSION

On these three angiosperms species, light compression wood was found in the opposite side of tension wood. Some more investigation, especially on the chemical composition and the anatomical structure, would be interesting to show whether this angiosperm compression wood differs or not from gymnosperm compression wood. In these trees, the re-orientation was produced by the cumulative effect of a high tension stress in the upper side of the tilted stem and a low compression stress in the lower side. Combining these two reactions seems to be the best way for optimising the biomechanical reactivity of the tree.

**Acknowledgements:** Authors thank Pr. Tanaka for his kindness in the use of X-Ray machine and G. Jaouen for shrinkage measurements on mani. Experiments on mani were performed in the laboratory "Ecologie des Forêts de Guyane" (UMR CIRAD – ENGREF – INRA 745– CNRS 2728) in Kourou (French Guyana) under the supervision of Pr. Meriem Fournier. The study was supported by the ADEME and the French Ministry of Agriculture in the framework of a project on "Physical and mechanical properties of reaction woods" (convention 61.45.47/00) and by a Grant in Aid for Scientific Research from the Japanese Society of Promotion of Science (no. 14656069, 14360099,14002805). First and second authors received post-doctoral fellowship from Japanese Society for Promotion of Sciences (JSPS).

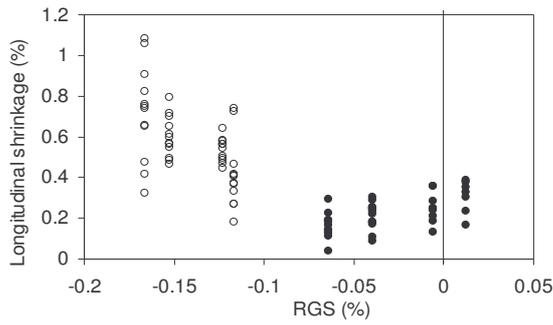

**Figure 1**: Longitudinal shrinkage measured on **chestnut** samples versus the residual growth strain. Full circle: normal and "compression" wood. Empty circle: tension wood (modified from [11]).

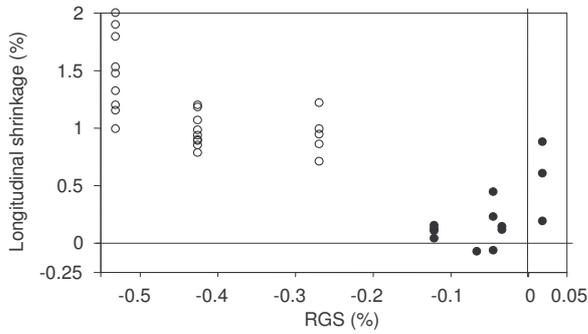

**Figure 2**: Longitudinal shrinkage measured on **mani** disks versus the approximate residual growth strain. Full circle: normal and "compression" wood. Empty circle: tension wood.

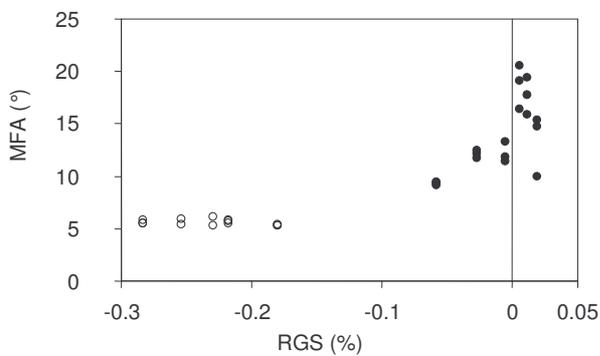

**Figure 3**: Microfibrils angle estimated on **poplar** samples versus the residual growth strain. Full circle: normal and "compression" wood. Empty circle: tension wood.